# Parallel Markov chain Monte Carlo for Bayesian hierarchical models with big data, in two stages

Zheng Wei[a] and Erin M. Conlon[b*]

[a]Department of Mathematics and Statistics, University of Maine, Orono, United States of America

[b]Department of Mathematics and Statistics, University of Massachusetts, Amherst, United States of America

*Erin M. Conlon, corresponding author; email: econlon@mathstat.umass.edu.

**Abstract**

Due to the escalating growth of big data sets in recent years, new Bayesian Markov chain Monte Carlo (MCMC) parallel computing methods have been developed. These methods partition large data sets by observations into subsets. However, for Bayesian nested hierarchical models, typically only a few parameters are common for the full data set, with most parameters being group-specific. Thus, parallel Bayesian MCMC methods that take into account the structure of the model and split the full data set by groups rather than by observations are a more natural approach for analysis. Here, we adapt and extend a recently introduced two-stage Bayesian hierarchical modeling approach, and we partition complete data sets by groups. In stage 1, the group-specific parameters are estimated independently in parallel. The stage 1 posteriors are used as proposal distributions in stage 2, where the target distribution is the full model. Using three-level and four-level models, we show in both simulation and real data studies that results of our method agree closely with the full data analysis, with greatly increased MCMC efficiency and greatly reduced computation times. The advantages of our method versus existing parallel MCMC computing methods are also described.

Keywords: Bayesian nested hierarchical models, parallel computing, big data, Markov chain Monte Carlo, Metropolis-Hastings algorithm, two stages

## 1. Introduction

With the rapid growth of big data, new scalable statistical and computational methods are needed for the analysis of large data sets. Here, big data refers to data sets that are too large to evaluate in full, due to memory constraints, limited storage capacity or excessive computation time. Several recent Bayesian and Markov chain Monte Carlo (MCMC) methods for large data sets have been introduced to address these issues. However, these methods either required communication between machines at each MCMC iteration, which slows computation times (Newman *et al*. [29]; Langford *et al*. [17]; Smola and Narayanamurthy [38]; Agarwal and Duchi [1]), or did not address issues with limited memory or limited storage capacity (Maclaurin and Adams [22]; Bardenet *et al*. [4]; Ahn *et al*. [2]; Quiroz *et al*. [35]; Chen *et al*. [8]; Wilkinson [43]; Laskey and Myers [18]; Murray [26]).

Another research direction involves dividing full data sets into subsets, performing independent Bayesian MCMC computation for each subset, and combining the independent results (Neiswanger *et al*. [28]; Scott *et al*. [38]). These are communication-free parallel methods, in that each machine generates MCMC samples without trading information with other machines. These techniques focus on data sets with a large number of observations relative to the dimensionality of the model, referred to as *tall data*, and they partition the data sets by observations into subsets.

The methods of Neiswanger *et al*. [28] and Scott *et al*. [38] have different procedures for combining the subset posterior samples. Neiswanger *et al*. [28] developed a kernel density estimator that evaluates each subset posterior density separately; these subset posteriors are then multiplied together to estimate the full data posterior. Alternatively, Scott *et al*. [38] created the consensus Monte Carlo algorithm that estimates the full data posterior through weighted averaging of the subset MCMC samples. Both approaches show good performance when the subset posteriors are near Gaussian, which is expected for adequately large sample sizes for each subset, based on the Bayesian central limit theorem (Bernstein von-Mises theorem; see Van der Vaart [41], and Le Cam and Yang [19]). However, for non-Gaussian posteriors, the methods may have unreliable performance (Baker *et al*. [3]; Neiswanger *et al*. [28]; Miroshnikov *et al*. [25]). The method of Neiswanger *et al*. [28] also has limitations as the number of unknown model parameters increases, since kernel density estimation becomes infeasible in larger dimensions (Wang and Dunson [42]; Scott [37]). Specifically, Scott [37] shows that kernel density estimation can break down for a model with as few as five to six parameters. Another drawback to the subset combining methods of Neiswanger *et al*. [28] and Scott *et al*. [38] is that the prior distribution for the full data analysis is split into $M$ components for the subset analyses, where $M$ is the number of subsets. As a result, the priors that are proper for the full data analysis can be improper for the subsets, leading to improper subset posteriors (Scott [37]).

Due to the limitations of the methods of Neiswanger *et al*. [28] and Scott *et al*. [38], we develop a two-stage communication-free parallel MCMC algorithm for Bayesian nested hierarchical models. The two-stage technique was introduced for Bayesian meta-analysis by Lunn *et al*. [20] and further modified and implemented by Bryan *et al*. [6], Hooten *et al*. [16], Peng *et al*. [31] and Goudie *et al*. [14], but was not carried out for parallel computing and big data. The motivation for the two-stage method is that for nested hierarchical models, generally only a few parameters are common for the full data set, with most parameters being group-specific. Thus, parallel MCMC methods that take into account the structure of the model and split the full data set by groups rather than by observations are a more natural procedure for these models (Scott *et al*. [38]). Our two-stage method takes this approach and partitions the full data analysis by groups; complete details are provided in Section 2. As a brief overview, in stage 1, the group-specific parameters are estimated in parallel, independently of other groups. The resulting posteriors are used as proposal distributions in a Metropolis-Hastings step in stage 2, where the target distribution is the full hierarchical model. The likelihood is not evaluated during stage 2 and therefore the full data set is never processed in its entirety; this speeds computation and allows for large data sets to be evaluated in parts.

The two-stage method has several advantages over existing communication-free parallel computing methods, including Neiswanger *et al*. [28] and Scott *et al*. [38]. Our method is appropriate for both Gaussian and non-Gaussian distributions, and the dimension of the parameter space is not a limitation. In addition, the priors for the full data analysis are not split into parts for the subset analyses, which avoids any improper subset priors. We illustrate these advantages in Sections 2 and 3.

**2. Methods**

To introduce the two-stage method, we first consider the following general Bayesian nested hierarchical model with three levels. For models with four or more levels, refer to Appendix 1.

$$\text{level 1:} \quad y_{ij} \mid \theta_i, \sigma_i^2 \sim p(y_{ij} \mid \theta_i, \sigma_i^2), \quad i=1,\ldots,n; \ j=1,\ldots,m_i,$$

$$\text{level 2:} \quad \theta_i \mid \mu, \tau^2 \sim p(\theta_i \mid \mu, \tau^2),$$

$$\sigma_i^2 \sim p(\sigma_i^2), \quad (1)$$

$$\text{level 3:} \quad \mu \sim p(\mu),$$

$$\tau^2 \sim p(\tau^2).$$

Here, the index $i$ denotes the group, and $y_{ij}$ is the data value $j$ from group $i$; the model parameters $\theta_i$, $\sigma_i^2$, $\mu$ and $\tau^2$ are unknown.

## 2.1. Full data analysis

For a full data analysis of the model in Equation (1), the likelihood of the data is

$$p(\boldsymbol{y} \mid \boldsymbol{\theta}, \boldsymbol{\sigma^2}) = \prod_{i=1}^{n} \prod_{j=1}^{m_i} p(y_{ij} \mid \theta_i, \sigma_i^2), \quad (2)$$

where $\boldsymbol{y}$ denotes the collection of all $y_{ij}s$, and $\boldsymbol{\theta}$ and $\boldsymbol{\sigma^2}$ denote the collection of all $\theta_i s$ and $\sigma_i^2 s$, respectively. The joint posterior distribution is

$$p(\boldsymbol{\theta}, \boldsymbol{\sigma^2}, \mu, \tau^2 \mid \boldsymbol{y}) \propto \left\{ \prod_{i=1}^{n} \prod_{j=1}^{m_i} p(y_{ij} \mid \theta_i, \sigma_i^2) \right\} \left\{ \prod_{i=1}^{n} p(\theta_i \mid \mu, \tau^2) p(\sigma_i^2) \right\} p(\mu) p(\tau^2). \quad (3)$$

In the traditional Bayesian analysis, the full model in Equation (1) is implemented for the entire data set, with all parameters estimated simultaneously. However, for instances where a full data analysis is not feasible, the following two-stage method is introduced.

## 2.2. Two-stage method

The two-stage method, first introduced by Lunn *et al*. [20], provides inferences on the full hierarchical model in Equation (1) without requiring evaluation of the complete data set in its entirety; details are described below.

### 2.2.1. Stage 1

In stage 1, the full data set is split by group $i$ and each group is analyzed independently in parallel. Here, independent prior distributions are assigned to each $\theta_i$. The specific choices of priors for the $\theta_i s$ are discussed in Section 3; these are typically chosen to be uninformative (Lunn *et al.* [20]). The common model parameters $\mu$ and $\tau^2$ are not estimated at this stage. For all remaining parameters, the priors are the same as in the full model. The stage 1 model is then:

$$\begin{aligned}
\text{level 1:} \quad & y_{ij} \mid \theta_i, \sigma_i^2 \sim p(y_{ij} \mid \theta_i, \sigma_i^2), \quad i=1,\dots,n;\ j=1,\dots,m_i, \\
\text{level 2:} \quad & \theta_i \sim p(\theta_i), \\
& \sigma_i^2 \sim p(\sigma_i^2).
\end{aligned} \qquad (4)$$

Here, we use the notation $\boldsymbol{y}_i$ to indicate the group-specific data values $y_{ij}$, $j=1,\dots,m_i$. MCMC samples are obtained independently for each group $i$, from the following joint posterior distribution of $\theta_i$ and $\sigma_i^2$, conditional on the subset data $\boldsymbol{y}_i$ only:

$$p(\theta_i, \sigma_i^2 \mid \boldsymbol{y}_i) \propto p(\boldsymbol{y}_i \mid \theta_i, \sigma_i^2) p(\theta_i) p(\sigma_i^2), \quad i=1,\dots,n. \qquad (5)$$

The MCMC sample size is $A_i$ for each group $i$, and the resulting MCMC samples are labeled $\{\theta_i^{(s)}, \sigma_i^{2(s)}\}$, $s=1,\dots,A_i$. The stage 1 samples are used as proposal distributions in stage 2 in a Metropolis-Hastings step that has the target distribution of the full model in Equation (1) (Metropolis *et al.* [24]; Hastings [15]).

*2.2.2. Stage 2*

Stage 2 uses a Metropolis-Hastings-within-Gibbs sampling scheme to iteratively draw from the joint posterior distribution of $\mu$, $\tau^2$, $\boldsymbol{\theta}$, and $\boldsymbol{\sigma}^2$ based on the full hierarchical model in Equation (1). For $T$ total samples in stage 2, at each iteration $t$, $t = 1,\dots,T$, we cycle through the full conditional posterior distributions of $\mu$ and $\tau^2$, and then jointly through each $\theta_i$ and $\sigma_i^2$, for $i = 1,\dots,n$. The full conditional posteriors are given by

$$p(\mu \mid \tau^2, \boldsymbol{\theta}, \boldsymbol{y}) \propto p(\mu) \prod_{i=1}^{n} p(\theta_i \mid \mu, \tau^2), \qquad (6)$$

$$p(\tau^2 \mid \mu, \boldsymbol{\theta}, \mathbf{y}) \propto p(\tau^2) \prod_{i=1}^{n} p(\theta_i \mid \mu, \tau^2), \tag{7}$$

$$p(\theta_i, \sigma_i^2 \mid \mu, \tau^2, \mathbf{y}) \propto p(\mathbf{y}_i \mid \theta_i, \sigma_i^2) p(\theta_i \mid \mu, \tau^2) p(\sigma_i^2), \quad i = 1, \ldots, n. \tag{8}$$

Initial values are assigned for $\tau^{2(0)}$, $\boldsymbol{\theta}^{(0)}$, and $\boldsymbol{\sigma}^{2(0)}$. The full conditional posterior distributions for the common model parameters $\mu$ and $\tau^2$ in Equations (6) and (7) are typically available in closed form and can then be sampled from directly using standard algorithms such as the Gibbs sampler (Gelfand and Smith [11]); otherwise, there are many alternative algorithms for sampling from these full conditional posteriors (e.g. Metropolis *et al.* [24]; Hastings [15]; Neal [27]; Gilks and Wild [13]).

For jointly updating each $\theta_i$ and $\sigma_i^2$ in Equation (8) at iteration $t$, a Metropolis-Hastings step is carried out. For this, the proposal distribution is based on stage 1 samples, and the target distribution is the full hierarchical model. Next, we introduce general notation for a Metropolis-Hastings step, and follow with the specific Metropolis-Hastings algorithm for our model.

*2.2.3. General Metropolis-Hastings algorithm*

For a random variable $\kappa$ that has density $h(\kappa)$, the Metropolis-Hastings algorithm produces dependent draws from $h(\kappa)$ by generating the Markov chain

$$\kappa^{(t)} = \begin{cases} \kappa^{*(t)} & \text{with probability} = \text{minimum}(1, r), \\ \kappa^{(t-1)} & \text{otherwise,} \end{cases} \tag{9}$$

for $t = 1, \ldots, T$. Here, $\kappa^{*(t)}$ is a proposal value sampled from a candidate density $q(\kappa)$ that approximates the target density $h(\kappa)$ reasonably well, but is easy to draw samples from. The value $r$ is the "acceptance probability" for the proposal value $\kappa^{*(t)}$, and is based on both target densities and proposal densities; $r$ is defined as

$$r = \frac{h(\kappa^{*(t)})}{h(\kappa^{(t-1)})} \frac{q(\kappa^{(t-1)})}{q(\kappa^{*(t)})}. \tag{10}$$

The value $r$ can be re-expressed as

$$r = \frac{h(\kappa^{*(t)})}{q(\kappa^{*(t)})} \frac{q(\kappa^{(t-1)})}{h(\kappa^{(t-1)})} = \frac{R(\kappa^{*(t)})}{R(\kappa^{(t-1)})}, \tag{11}$$

where $R(x)$ denotes the target-to-candidate ratio of densities $h(x)/q(x)$ (see also Lunn et al. [20]).

*2.2.4. Metropolis-Hastings step for stage 2*

For our model, we use a Metropolis-Hastings step at iteration $t$ to jointly update $\theta_i$ and $\sigma_i^2$ based on Equation (8). For this, we sample uniformly at random a candidate value $\{\theta_i^{*(t)}, \sigma_i^{2*(t)}\}$ from the $A_i$ samples from stage 1 and label it $a_{it}$. This candidate value $\{\theta_i^{*(t)}, \sigma_i^{2*(t)}\}$ has the following posterior density from stage 1 in Equation (5), which we label $q(\theta_i, \sigma_i^2)$; note that this density contains independent priors for the $\theta_i s$:

$$\{\theta_i^{*(t)}, \sigma_i^{2*(t)}\} = \{\theta_i^{(a_{it})}, \sigma_i^{2(a_{it})}\} \sim p(\mathbf{y}_i | \theta_i, \sigma_i^2) p(\theta_i) p(\sigma_i^2) = q(\theta_i, \sigma_i^2). \tag{12}$$

In determining the ratio $R$ for this candidate value $\{\theta_i^{*(t)}, \sigma_i^{2*(t)}\}$, the target posterior density of the full model in Equation (8) is also evaluated at the candidate value $\{\theta_i^{*(t)}, \sigma_i^{2*(t)}\}$. We label this target posterior density as $h(\theta_i, \sigma_i^2)$; note that it includes priors for $\theta_i$ that are dependent on $\mu$ and $\tau^2$, as follows:

$$p(\theta_i, \sigma_i^2 | \mu, \tau^2, \mathbf{y}) \propto p(\mathbf{y}_i | \theta_i, \sigma_i^2) p(\theta_i | \mu, \tau^2) p(\sigma_i^2) = h(\theta_i, \sigma_i^2). \tag{13}$$

The target-to-candidate ratio for $\{\theta_i^{*(t)}, \sigma_i^{2*(t)}\}$ is then

$$\begin{aligned} R(\theta_i^{*(t)}, \sigma_i^{2*(t)}) &= \frac{h(\theta_i^{*(t)}, \sigma_i^{2*(t)})}{q(\theta_i^{*(t)}, \sigma_i^{2*(t)})} \propto \frac{p(\mathbf{y}_i | \theta_i^{(a_{it})}) p(\theta_i^{(a_{it})} | \mu^{(t)}, \sigma^{2(t)}) p(\sigma_i^{2(a_{it})})}{p(\mathbf{y}_i | \theta_i^{(a_{it})}) p(\theta_i^{(a_{it})}) p(\sigma_i^{2(a_{it})})} \\ &= \frac{p(\theta_i^{(a_{it})} | \mu^{(t)}, \tau^{2(t)})}{p(\theta_i^{(a_{it})})}. \end{aligned} \tag{14}$$

Similarly, the target-to-candidate ratio for the previous value of the Markov chain, $\{\theta_i^{(t-1)}, \sigma_i^{2(t-1)}\}$, is

$$R(\theta_i^{(t-1)}, \sigma_i^{2(t-1)}) = \frac{h(\theta_i^{(t-1)}, \sigma_i^{2(t-1)})}{q(\theta_i^{(t-1)}, \sigma_i^{2(t-1)})} \propto \frac{p(\mathbf{y}_i \mid \theta_i^{(t-1)}) p(\theta_i^{(t-1)} \mid \mu^{(t)}, \tau^{2(t)}) p(\sigma_i^{2(t-1)})}{p(\mathbf{y}_i \mid \theta_i^{(t-1)}) p(\theta_i^{(t-1)}) p(\sigma_i^{2(t-1)})}$$
$$= \frac{p(\theta_i^{(t-1)} \mid \mu^{(t)}, \tau^{2(t)})}{p(\theta_i^{(t-1)})}.$$
(15)

Thus, $r$ is given by the following, and the candidate value $\{\theta_i^{*(t)}, \sigma_i^{2*(t)}\}$ will be accepted with the probability of minimum$(1, r)$:

$$r = \frac{h(\theta_i^{*(t)}, \sigma_i^{2*(t)})}{q(\theta_i^{*(t)}, \sigma_i^{2*(t)})} \frac{q(\theta_i^{(t-1)}, \sigma_i^{2(t-1)})}{h(\theta_i^{(t-1)}, \sigma_i^{2(t-1)})} = \frac{R(\theta_i^{*(t)}, \sigma_i^{2*(t)})}{R(\theta_i^{(t-1)}, \sigma_i^{2(t-1)})} = \frac{\dfrac{p(\theta_i^{(a_{it})} \mid \mu^{(t)}, \tau^{2(t)})}{p(\theta_i^{(a_{it})})}}{\dfrac{p(\theta_i^{(t-1)} \mid \mu^{(t)}, \tau^{2(t)})}{p(\theta_i^{(t-1)})}}$$
$$= \frac{p(\theta_i^{(a_{it})} \mid \mu^{(t)}, \tau^{2(t)})}{p(\theta_i^{(t-1)} \mid \mu^{(t)}, \tau^{2(t)})} \frac{p(\theta_i^{(t-1)})}{p(\theta_i^{(a_{it})})}.$$
(16)

Note that the likelihoods cancel in Equations (14) and (15), and are thus not included in $r$. As a result, the data sets are not analyzed in stage 2, and this stage is computationally very fast. If the stage 1 prior for $\theta_i$ is effectively uniform, then $r$ can be simplified further as

$$r = \frac{p(\theta_i^{(a_{it})} \mid \mu^{(t)}, \tau^{2(t)})}{p(\theta_i^{(t-1)} \mid \mu^{(t)}, \tau^{2(t)})}.$$
(17)

The value $r$ in Equation (17) is the ratio of priors for $\theta_i$ from the full model, evaluated at the candidate value for $\theta_i$ in the numerator and the previous value of the Markov chain for $\theta_i$ in the denominator. Note that $r$ does not depend on $\sigma_i^2$ in both Equations (16) and (17), so that these values are not used in computations in stage 2. Convergence of the MCMC sampler in stage 2 is assessed using standard methods (Cowles and Carlin [9]; Mengersen *et al*. [23]); we detail convergence diagnostics in Section 2.3.

### 2.3. Computational details and convergence diagnostics

For the examples in Section 3, we used the R programming language (R Core Team [36]) for all computations. For stage 1 and for the full data model, the R package RJAGS [34] is used to run the JAGS program (Plummer [32]) to generate the MCMC samples; other standard MCMC software such as Stan (Carpenter *et al*. [7]) or OpenBUGS (Lunn *et al*. [21]) can also be used for stage 1. For all examples, we chose data set sizes so that a full data analysis was still feasible; this was in order to compare the two-stage results and the full data results.

For stage 1, stage 2 and the full data analyses in all examples, we ran two parallel MCMC chains with MCMC sample size of 250,000 after convergence. Initial values of parameters were overdispersed and widely different for the two chains. The resulting samples were thinned by 10 (see Lunn *et al*. [20]) based on auto-correlation values; this produced samples that were effectively independent between consecutive values, with a final MCMC sample size of 50,000 for the two chains. Convergence was assessed using history plots and Gelman and Rubin statistics (Gelman and Rubin [12]; Brooks and Gelman [5]), which also showed that a burnin of 10,000 was acceptable for all analyses (Cowles and Carlin [9]; Mengersen *et al*. [23]). The MCMC sample sizes were equivalent for the two-stage method and the full data analyses, for comparison purposes.

## 2.4. MCMC efficiency and computation times

We compare the results for the two-stage method and the full data analysis using minimum MCMC efficiency and computation time measured in both CPU and elapsed time; all measures are compared after burnin. Estimated relative $L_1$ and $L_2$ distances are also used for comparisons (Appendix 2). We use the following definition for minimum MCMC efficiency:

$$\text{Minimum MCMC efficiency} = \min\left(\frac{\text{Effective sample size}}{\text{CPU computation time in seconds}}\right) \quad (18)$$

Here, effective sample size is based on the CODA R package (Plummer *et al*. [33]); we use the minimum efficiency over all groups and all parameters for each analysis. Since the two-stage method requires twice as many MCMC samples as the full data method over the two stages, we created the MCMC efficiency improvement factor. This is calculated as follows, assuming equal MCMC sample sizes in all stages and analyses:

MCMC efficiency improvement factor, two stage versus full data analysis =
$$\left(\frac{\text{Average minimum MCMC efficiency over two stages in two-stage method}}{\text{Minimum MCMC efficiency in full data analysis}}\right) \times 0.5. \quad (19)$$

## 3. Examples

### 3.1. Simulation study 1: three-level hierarchical normal model

Here, we simulate data for the following three-level Bayesian nested hierarchical normal model:

$$
\begin{aligned}
&\text{level 1: } y_{ij} \mid \theta_i, \sigma_i^2 \sim Normal(\theta_i, \sigma_i^2), \ i=1,\ldots,n; \ j=1,\ldots,m_i, \\
&\text{level 2: } \theta_i \mid \mu, \tau^2 \sim Normal(\mu, \tau^2), \\
&\quad\quad\quad\ \sigma_i^2 \sim Inverse\ Gamma(0.01, 0.01), \\
&\text{level 3: } \mu \sim Normal(0, 10^6), \\
&\quad\quad\quad\ \tau^2 \sim Inverse\ Gamma(0.1, 0.1).
\end{aligned}
\tag{20}
$$

We produced three data sets, one for each of the number of groups $n = 20, 50, 100$. For each data set, we assigned $\mu = 25$, $\tau^2 = 1.5$; the $\theta_i$ values were simulated from Model (20) based on these values, and the $\sigma_i^2$ were simulated from a normal distribution with mean 10 and variance 1. A sample size of $m_i = 100{,}000$ per group was generated for each data set.

#### 3.1.1. Two stage method

##### 3.1.1.1. Stage 1.
Here, we split the analysis at level 2, so that each group $i$ is analyzed independently in parallel, with the $\theta_i s$ assigned vague independent priors (Lunn *et al*. [20]). The $\sigma_i^2 s$ are assigned the same priors as in the full model in Equation (20). This results in the following stage 1 model:

$$
\begin{aligned}
&\text{level 1: } y_{ij} \mid \theta_i, \sigma_i^2 \sim Normal(\theta_i, \sigma_i^2), \ i=1,\ldots,n; \ j=1,\ldots,m_i, \\
&\text{level 2: } \theta_i \sim Normal(0, 10^6), \\
&\quad\quad\quad\ \sigma_i^2 \sim Inverse\ Gamma(0.01, 0.01).
\end{aligned}
\tag{21}
$$

The common model parameters $\mu$ and $\tau^2$ are not estimated in stage 1. We use the approaches detailed in Sections 2.2 and 2.3 to produce $A_i = 50{,}000$ samples from the joint posterior distribution of $\theta_i$ and $\sigma_i^2$ conditional on the subset data $y_i$ for each group independently in parallel. The resulting sampled values are used as proposal distributions in stage 2.

*3.1.1.2. Stage 2.* For stage 2, we estimate the full model in Equation (20) by cycling through the Metropolis-Hastings-within-Gibbs algorithm for all unknown parameters $\mu, \tau^2, \boldsymbol{\theta},$ and $\boldsymbol{\sigma}^2$, as detailed in Section 2.2. An MCMC sample size of $T = 50,000$ is again produced as described in Section 2.3. The full conditional posterior distributions are closed form for $\mu$ and $\tau^2$, so that a Gibbs sampling algorithm is carried out for these parameters.

*3.1.2. Simulation study 1 results*

As stated in Section 2.3, for the full data analysis, we used the same MCMC sample size of $T = 50,000$ that was used in the two-stage method for comparison (see also Lunn *et al.* [20]). For $n = 50$ groups, the MCMC efficiency improvement factor for the two-stage method versus the full data method was 27.8, indicating a 27.8-fold improvement in efficiency for the two-stage method (Table 1). Both the CPU and elapsed computation times were reduced by 96.7% for the two-stage method versus the full data analysis, assuming all groups were run in parallel. Note that these time reductions are dependent on the number of groups (Table 2). We find a close agreement between results of our two-stage method and the full data approach based on the estimated relative $L_1$ and $L_2$ distances (Appendix 2); the average estimated relative $L_1$ and $L_2$ distances are between 0.021 and 0.024 for all parameters (Table 3). For $n = 20$ and 100 groups, the results are similar to those for $n = 50$ (Tables 1, 2, and 3).

Figure 1 shows the marginal posterior distributions for representative parameters for both the full data analysis and the two-stage method for $n = 50$ groups; these plots illustrate the close similarity of our two-stage method to the complete data analysis. Note that the variance parameter $\tau^2$ has a skewed posterior distribution, and the estimated relative $L_1$ and $L_2$ distances are 0.023 and 0.024, respectively; this indicates that our method is appropriate for non-Gaussian as well as Gaussian posteriors. We also plot the stage 1 results for representative parameters in Figure 1. For the two-stage method and large data sets, the stage 1 and stage 2 posteriors are similar. This is due to the individual groups having large data sizes, which result in weak shrinkage effects.

### 3.2. Simulation study 2: four-level hierarchical logistic regression model

Here, we simulate data for the following four-level Bayesian hierarchical logistic regression model:

$$\begin{aligned}
&\text{level 1: } y_{ij} \sim Bernoulli(p_{ij}),\ i=1,...,n;\ j=1,...,m_i, \\
&\text{level 2: } logit(p_{ij}) = X_{ij}^T \boldsymbol{\beta}_i, \\
&\text{level 3: } \boldsymbol{\beta}_i \sim Normal_k(\boldsymbol{\mu}, \boldsymbol{\Sigma}), \\
&\text{level 4: } \mu_k \sim Normal(0, \tau_k^2),\ k=1,...,K, \\
&\boldsymbol{\Sigma}^{-1} \sim Wishart_k(\boldsymbol{\Psi}, \nu),
\end{aligned} \quad (22)$$

where $p_{ij} = \Pr(y_{ij} = 1)$, $logit(p_{ij}) = \log(p_{ij}/(1-p_{ij}))$, $n$ = number of groups, $m_i$ = sample size per group, and $K$ = number of predictor variables. Here, $\boldsymbol{\beta}_i$ is a $K$-dimensional vector for group $i$ consisting of $\beta_{i1},...,\beta_{iK}$; $X_{ij}$ represents the $m_i$ measurements in group $i$ for the $K$ predictor variables, $\boldsymbol{\mu}$ is a $K$-dimensional vector, $\boldsymbol{\Sigma}$ and $\boldsymbol{\Psi}$ are $K \times K$-dimensional matrices, and $\nu$ is the shape parameter of the Wishart distribution. We use $n = 20$, $m_i = 10{,}000$, $K = 3$, and assigned $\boldsymbol{\mu} = (0.8, -0.9, 0.7)'$. For $\boldsymbol{\Sigma}$, we specified $\Sigma_{kk'} = 0.05$ for $k=k'$, and $\Sigma_{kk'} = 0.04$ for $k \neq k'$. We assigned $\boldsymbol{\Psi}$ to be a diagonal matrix with diagonal elements $\Psi_{kk} = 0.01$, $k=1,...,K$; we also specified $\tau_k^2 = 10$, $k = 1,...,K$; and $\nu = 4$. The predictor values were simulated as $X_{ij} \sim Normal(\boldsymbol{0}, \Lambda)$, where $\Lambda_{kk'} = 1$ for $k=k'$, and $\Lambda_{kk'} = 0.2$ for $k \neq k'$. With these specifications, we simulated the $\boldsymbol{\beta}_i$, $p_{ij}$ and $y_{ij}$ values based on Model (22).

#### 3.2.1. Two stage method

A description of the two-stage approach for a four-level hierarchical model are provided in Appendix 1; we summarize the procedure next.

*3.2.1.1. Stage 1.* For the full model in Equation (22), we divide the computations at level 3, so that each group $i$ is evaluated independently in parallel, with the $\boldsymbol{\beta}_i$s assigned vague independent priors (Lunn *et al*. [20]). The common model parameters $\boldsymbol{\mu}$ and $\boldsymbol{\Sigma}$ are not estimated in stage 1. This results in the following stage 1 model:

$$\text{level 1: } y_{ij} \sim Bernoulli(p_{ij}), \ i=1,...,n; \ j=1,...,m_i,$$
$$\text{level 2: } logit(p_{ij}) = X_{ij}^T \boldsymbol{\beta}_i, \quad (23)$$
$$\text{level 3: } \boldsymbol{\beta}_i \sim Normal_k(\boldsymbol{0}, \boldsymbol{\Omega}),$$

where $\boldsymbol{\Omega}$ is a diagonal matrix with diagonal elements $\Omega_{kk} = 100, \ k = 1,...,K$. We again use the methods described in Appendix 1 and Section 2.3 to create $A_i$ = 50,000 samples from the posterior distribution of $\boldsymbol{\beta}_i$ conditional on the subset data $y_i$ independently in parallel for each group. The posterior samples from stage 1 are used as proposal distributions in stage 2.

*3.2.1.2. Stage 2.* We estimate the full model in Equation (22) in stage 2 by implementing the Metropolis-Hastings-within-Gibbs algorithm for all unknown parameters $\boldsymbol{\mu}, \boldsymbol{\Sigma}, \boldsymbol{\beta}_i$, $i = 1,…,n$, as detailed in Appendix 1. An MCMC sample size of $T$ = 50,000 is again produced as described in Section 2.3. The full conditional posterior distributions are closed form for $\boldsymbol{\mu}$ and $\boldsymbol{\Sigma}$, so that a Gibbs sampling procedure is used for these parameters.

*3.2.2. Simulation study 2 results*

For comparison purposes, we used the same MCMC sample size of $T$ = 50,000 for the full data analysis and the computations in stages 1 and 2 (see also Lunn *et al*. [20]); Section 2.3 provides details, including convergence diagnostics. For all results, our findings are similar to simulation study 1. In particular, the MCMC efficiency improvement factor for the two-stage method compared to the full data method was 21.9, resulting in a 21.9-fold improvement in efficiency for the two-stage technique (Table 1). Both the CPU and elapsed computation times were reduced by 89.0% for the two-stage versus full data approaches, assuming that all groups are processed in parallel; these decreases in computation times depend on the number of groups (Table 2). The estimated relative $L_1$ and $L_2$ distances range from 0.018 and 0.025 for all model parameters (Table 3).

*3.3. Real airlines data and four-level model*

Here, we examine real data for all commercial flights within the United States for the twelve-month period August 2016 to July 2017 (U. S. Department of Transportation [40]). The outcome variable of interest is the arrival delay for each flight, measured in minutes; flights with arrival delay of fifteen minutes or less are regarded as on-time [40]. There were a total of 1,061,023 data values for arrival delays. A log-transformation was applied to the data set, and the resulting transformed data values were approximately normally distributed. Data was available for twelve airlines and seven days of the week. We use the following four-level Bayesian nested hierarchical model to analyze the full data set:

$$\begin{aligned}
&\text{level 1: } y_{ijk} \mid \delta_{ij}, \eta_{ij}^2 \sim Normal(\delta_{ij}, \eta_{ij}^2), \ i=1,\ldots,n; \ j=1,\ldots,m; \ k=1,\ldots,K_{ij}, \\
&\text{level 2: } \delta_{ij} \mid \theta_i, \sigma_i^2 \sim Normal(\theta_i, \sigma_i^2), \\
&\quad\quad\quad \eta_{ij}^2 \sim Inverse\ Gamma(0.1, 0.1), \\
&\text{level 3: } \theta_i \mid \mu, \tau^2 \sim Normal(\mu, \tau^2), \\
&\quad\quad\quad \sigma_i^2 \sim Inverse\ Gamma(0.01, 0.01), \\
&\text{level 4: } \mu \sim Normal(0, 10^6), \\
&\quad\quad\quad \tau^2 \sim Inverse\ Gamma(0.1, 0.1).
\end{aligned} \quad (24)$$

Here, the $y_{ijk}$ are the log-transformed arrival delays for airline $i$, $i = 1,\ldots,n$; day of the week $j$, $j = 1,\ldots,m$; and data replicate $k$ within carrier $i$ and day $j$, $k = 1,\ldots,K_{ij}$. The $\delta_{ij}$ parameters are the mean arrival delays for each airline $i$ and day $j$; $\theta_i$ are the means for each airline $i$, and $\mu$ is the mean over all airlines. The unknown variance parameters are the variability within airline $i$ and day $j$, $\eta_{ij}^2$; the variability with airline $i$, $\sigma_i^2$; and the variability over all airlines, $\tau^2$.

*3.3.1. Two-stage method*

Complete details of the two-stage approach for a general four-level nested hierarchical model are provided in Appendix 3, with summarized steps given below.

*3.3.1.1. Stage 1.* For the full model in Equation (24), we split the analysis at level 3, so that data values within group $i$ (i.e. airline $i$) are analyzed in parallel independently in stage 1. The $\theta_i s$ are assigned independent vague priors, with all other priors the same as in the full model (Lunn *et al*. [20]). The stage 1 model is then:

$$\begin{aligned}
&\text{level 1:} \quad y_{ijk} \mid \delta_{ij}, \eta_{ij}^2 \sim Normal(\delta_{ij}, \eta_{ij}^2), \quad i=1,\ldots,n;\ j=1,\ldots,m;\ k=1,\ldots,K_{ij}, \\
&\text{level 2:} \quad \delta_{ij} \mid \theta_i, \sigma_i^2 \sim Normal(\theta_i, \sigma_i^2), \\
&\phantom{\text{level 2:} \quad} \eta_{ij}^2 \sim Inverse\ Gamma(0.1, 0.1), \quad\quad\quad\quad\quad\quad\quad\quad\quad\quad (25) \\
&\text{level 3:} \quad \theta_i \sim Normal(0, 10^6), \\
&\phantom{\text{level 3:} \quad} \sigma_i^2 \sim Inverse\ Gamma(0.01, 0.01).
\end{aligned}$$

The common model parameters $\mu$ and $\tau^2$ are not estimated in stage 1. The notation $\mathbf{y}_{ij}$ is used to denote $y_{ijk}$, $k=1,\ldots,K_{ij}$; $\boldsymbol{\delta}_i$ denotes $\delta_{ij}$, $j=1,\ldots,m_i$; and $\boldsymbol{\eta}_i^2$ denotes $\eta_{ij}^2$, $j=1,\ldots,m_i$. For each airline $i$, we use the procedures described in Appendix 3 and Section 2.3 to generate $A_i$ = 50,000 samples independently in parallel from the joint distribution of $\boldsymbol{\delta}_i$, $\boldsymbol{\eta}_i^2$, $\theta_i$ and $\sigma_i^2$, conditional on the subset data $\mathbf{y}_{ij}$. The stage 1 samples are used as proposal distributions in stage 2.

*3.3.1.2. Stage 2.* For stage 2, the full model in Equation (24) is estimated by iterating through the Gibbs sampler for the common model parameters $\mu$ and $\tau^2$, which are closed form, and the Metropolis-Hastings step for each $\boldsymbol{\delta}_i$, $\boldsymbol{\eta}_i^2$, $\theta_i$ and $\sigma_i^2$ jointly, with an MCMC sample size of $T$ = 50,000. Full details of the implementation of stage 2 for a general four-level model are provided in Appendix 3 and Section 2.3.

### 3.3.2. Real data results

For the full data analysis, we used the same MCMC sample size of $T$ = 50,000 as in stages 1 and 2 for comparison (see also Lunn *et al*. [20]); see Section 2.3 for implementation details, including convergence diagnostics. For all results, our findings are similar to the simulation studies. Specifically, the MCMC efficiency improvement factor for the two-stage method versus the full data method was 17.1, resulting in a 17.1-fold increase in efficiency for the two-stage approach (Table 1). The CPU time was decreased by 65.3% for the two-stage versus full data methods, and the elapsed time was lowered by 65.2%, assuming all airlines are computed in parallel; these time reductions are dependent on the number of airlines (Table 2). The estimated relative $L_1$ and $L_2$ distances range between 0.017 and 0.031 for all parameters (Table 3). Figure 2 illustrates the close agreement of our two-stage method with the results from the complete data analysis.

## 4. Conclusions

Here, we adapted and extended the two-stage Bayesian hierarchical modeling approach introduced by Lunn et al. [20] for communication-free parallel MCMC and big data. The two-stage method takes into account the structure of the model and partitions the full data set by groups rather than by observations. We found in simulation studies and a real data analysis that our two-stage method produces results that closely agree with the full data analysis, with greatly increased MCMC efficiencies and greatly reduced computation times. The increases in efficiencies and reductions in computation times are dependent on the number of groups.

The two-stage method has several advantages over the existing communication-free parallel computing methods of Neiswanger et al. [28] and Scott et al. [38] that divide the full data set by observations rather than by groups. As shown in the examples, our method is appropriate for both Gaussian and non-Gaussian posterior distributions, unlike these existing methods that are best suited to Gaussian posteriors. The two-stage method also does not split the prior distributions of the full model into parts for the subset analyses, avoiding any impropriety of subset priors and resulting subset posteriors. In addition, unlike the kernel density estimation technique of Neiswanger et al. [28], our procedure is not limited by the dimension of the number of groups. This is due to the groups being processed independently in stage 1, and then estimated one-by-one in stage 2 through the Metropolis-Hastings step. The two-stage method is also flexible in that there can be different numbers of MCMC samples drawn in each group in stage 1; this may be necessary if, for example, some groups require more iterations to converge than others. In addition, the parallel computing in stage 1 is particularly applicable for distributed computing frameworks such as MapReduce (Dean and Ghemawat [10]), and can be run on multi-core processors and networks of machines.

A limitation of the two-stage method is that the stage 1 posteriors must be reasonable approximations to the full data posteriors (Lunn et al. [20]). If there is too much difference between these posteriors, then the Metropolis-Hastings algorithm will have a very low acceptance rate. However, with large data sets, this is unlikely to happen, since there will typically be much more weight for the posteriors toward the individual groups and the shrinkage effects are expected to be weak. As a result, the stage 1 posteriors closely resemble the full data posteriors; this is shown in our first simulation example and the real data example (Figures 1 and 2).

**Disclosure statement**

No potential conflict of interest was reported by the authors.

**Appendix 1**

*Hierarchical models with more than three levels*

As stated in Lunn *et al*. [20], the two-stage approach can be revised directly for models with four levels or more. To summarize the approach outlined in Lunn *et al*. [20], for a model with $C$ levels, the analysis can be split at level $c_s$, $1 \leq c_s \leq C$. In stage 1, independent posteriors for the parameters of interest at level $c_s$ are sampled, and these samples are used as proposal distributions in stage 2. The two-stage method then proceeds similarly to that described in Section 2.2. The likelihood and additional parameters in levels 1 to $c_s$ will cancel in Equations (14) and (15), if the distributional assumptions and data values are the same in the stage 1 model and the full hierarchical model. All parameters in levels $(c_s + 1),...,C$ are estimated in stage 2 using either Gibbs sampling or alternative standard MCMC algorithms (Gelfand and Smith [11]; Metropolis *et al*. [24]; Hastings [15]; Neal [27]; Gilks and Wild [13]). Full details for the two-stage approach with a general four-level nested hierarchical model are provided in Appendix 3; we implement a simulation study for a four-level model in Section 3.2, and a real data analysis for a four-level model in Section 3.3.

**Appendix 2**

*Metrics for comparing posterior densities*

The performance of the two-stage method is assessed by comparing results to the full data analysis. For this, we measure the distance between the marginal posterior density $p^F$ based on the full data approach and the marginal posterior density $p^T$ based on the two-stage analysis. For parameter $\psi$, we measure estimated relative $L_1$ distance, $d_1(p^F, p^T)$, and estimated relative $L_2$ distance, $d_2(p^F, p^T)$, relative to the full data marginal posterior; these are defined as follows:

$$d_1(p^F, p^T) = \frac{\|p^F - p^T\|_{L_1}}{\|p^F\|_{L_1}} = \frac{\int_R |p^F(\psi) - p^T(\psi)| d\psi}{\int_R |p^F(\psi)| d\psi}, \text{ and} \quad (26)$$

$$d_2(p^F, p^T) = \frac{\|p^F - p^T\|_{L_2}}{\|p^F\|_{L_2}} = \frac{\left(\int_R (p^F(\psi) - p^T(\psi))^2 d\psi\right)^{1/2}}{\left(\int_R (p^F(\psi))^2 d\psi\right)^{1/2}}. \quad (27)$$

Density smoothing is used for $p^F$ and $p^T$, as described in Neiswanger *et al.* [28] and Oliva *et al.* [30]. Values of the estimated relative $L_1$ and $L_2$ distances near zero indicate a close agreement of the two-stage estimated posterior to the full data posterior. The estimated relative $L_1$ and $L_2$ distances are produced for all unknown model parameters in the examples in Section 3.

**Appendix 3**

*Two-stage method for a general four-level Bayesian nested hierarchical model*

Here, we describe the two-stage method for the following general four-level Bayesian nested hierarchical model:

$$\begin{aligned}
\text{level 1:} \quad & y_{ijk} \mid \delta_{ij}, \eta_{ij}^2 \sim p(y_{ijk} \mid \delta_{ij}, \eta_{ij}^2), \ i = 1, ..., n; \ j = 1, ..., m_i; \ k = 1, ..., K_{ij}, \\
\text{level 2:} \quad & \delta_{ij} \mid \theta_i, \sigma_i^2 \sim p(\delta_{ij} \mid \theta_i, \sigma_i^2), \\
& \eta_{ij}^2 \sim p(\eta_{ij}^2), \\
\text{level 3:} \quad & \theta_i \mid \mu, \tau^2 \sim p(\theta_i \mid \mu, \tau^2), \quad (28) \\
& \sigma_i^2 \sim p(\sigma_i^2), \\
\text{level 4:} \quad & \mu \sim p(\mu), \\
& \tau^2 \sim p(\tau^2).
\end{aligned}$$

*Stage 1*

In stage 1, the analysis is split by group *i* at level 3 and each group *i* is evaluated independently. The $\theta_i$ s are assigned independent prior distributions, and the $\mu$ and $\tau^2$ common model parameters are not estimated at this stage. The stage 1 model is then:

$$\begin{aligned}
\text{level 1:} \quad & y_{ijk} \mid \delta_{ij}, \eta_{ij}^2 \sim p(y_{ijk} \mid \delta_{ij}, \eta_{ij}^2), \ i=1,...,n; \ j=1,...,m_i; \ k=1,...,K_{ij}, \\
\text{level 2:} \quad & \delta_{ij} \mid \theta_i, \sigma_i^2 \sim p(\delta_{ij} \mid \theta_i, \sigma_i^2), \\
& \eta_{ij}^2 \sim p(\eta_{ij}^2), \\
\text{level 3:} \quad & \theta_i \sim p(\theta_i), \\
& \sigma_i^2 \sim p(\sigma_i^2).
\end{aligned} \quad (29)$$

Here, we use the notation $\boldsymbol{y}_{ij}$ to indicate $y_{ijk}$, $k=1,...,K_{ij}$; $\boldsymbol{\delta}_i$ for $\delta_{ij}$, $j=1,...,m_i$; and $\boldsymbol{\eta}_i^2$ for $\eta_{ij}^2$, $j=1,...,m_i$. MCMC samples are generated independently in parallel for each group $i$ from the following joint posterior distribution of $\boldsymbol{\delta}_i$, $\boldsymbol{\eta}_i^2$, $\theta_i$ and $\sigma_i^2$ conditional on the subset data $\boldsymbol{y}_{ij}$ only:

$$p(\boldsymbol{\delta}_i, \boldsymbol{\eta}_i^2, \theta_i, \sigma_i^2 \mid \boldsymbol{y}_{ij}) \propto p(\boldsymbol{y}_{ij} \mid \boldsymbol{\delta}_i, \boldsymbol{\eta}_i^2) p(\boldsymbol{\delta}_i \mid \theta_i, \sigma_i^2) p(\boldsymbol{\eta}_i^2) p(\theta_i) p(\sigma_i^2), \ i=1,...,n. \quad (30)$$

The resulting MCMC samples of size $A_i$ for each group $i$ are labeled $\{\boldsymbol{\delta}_i^{(s)}, \boldsymbol{\eta}_i^{2(s)}, \theta_i^{(s)}, \sigma_i^{2(s)}\}$, $s=1,...,A_i$, for $i=1,...,n$. The stage 1 posteriors are used as proposal distributions in a Metropolis-Hastings step in stage 2.

*Stage 2*

Stage 2 uses the same Metropolis-Hastings-within-Gibbs sampling scheme described in Section 2.2 to iteratively draw from the joint posterior distribution of $\mu, \tau^2, \boldsymbol{\delta}, \boldsymbol{\eta}^2, \boldsymbol{\theta}$ and $\boldsymbol{\sigma}^2$ based on the full hierarchical model in Equation (28), where $\boldsymbol{\delta}, \boldsymbol{\eta}^2, \boldsymbol{\theta}$ and $\boldsymbol{\sigma}^2$ denote the collection of all $\delta_{ij}$s, $\eta_{ij}^2$s, $\theta_i$s and $\sigma_i^2$s, respectively. Initial values $\tau^{2(0)}$, $\boldsymbol{\delta}^{(0)}$, $\boldsymbol{\eta}^{2(0)}$, $\boldsymbol{\theta}^{(0)}$ and $\boldsymbol{\sigma}^{2(0)}$ are assigned to the parameters in the full conditional posterior distributions, except for the first sampled parameter $\mu$, which is conditional on all other initial values. At each iteration $t$, $t = 1,...,T$, we cycle through the full conditional posterior distributions of $\mu$ and $\tau^2$, and then jointly through each set of parameters $\boldsymbol{\delta}_i$, $\boldsymbol{\eta}_i^2$, $\theta_i$ and $\sigma_i^2$ of group $i$, for $i = 1,...,n$. Specifically, from Model (28), the full conditional posterior distributions are the following:

$$p(\mu | \tau^2, \boldsymbol{\delta}, \boldsymbol{\eta}^2, \boldsymbol{\theta}, \boldsymbol{\sigma}^2, \boldsymbol{y}) \propto p(\mu) \prod_{i=1}^{n} p(\theta_i | \mu, \tau^2), \tag{31}$$

$$p(\tau^2 | \mu, \boldsymbol{\delta}, \boldsymbol{\eta}^2, \boldsymbol{\theta}, \boldsymbol{\sigma}^2, \boldsymbol{y}) \propto p(\tau^2) \prod_{i=1}^{n} p(\theta_i | \mu, \tau^2), \tag{32}$$

$$p(\boldsymbol{\delta}_i, \boldsymbol{\eta}_i^2, \theta_i, \sigma_i^2 | \mu, \tau^2, \boldsymbol{y}) \propto \\ p(\boldsymbol{y}_{ij} | \boldsymbol{\delta}_i, \boldsymbol{\eta}_i^2) p(\boldsymbol{\delta}_i | \theta_i, \sigma_i^2) p(\boldsymbol{\eta}_i^2) p(\theta_i | \mu, \tau^2) p(\sigma_i^2), \quad i=1,...,n, \tag{33}$$

where $\boldsymbol{y}$ denotes the collection of all $y_{ijk}s$. The full conditional distributions for the common model parameters $\mu$ and $\tau^2$ in Equations (31) and (32) are often available in closed form and can then be sampled from directly through algorithms such as the Gibbs sampler (Gelfand and Smith, 1990); otherwise, many alternative algorithms can be used to sample from these full conditional posterior distributions (e.g. Metropolis *et al.*, 1953; Hastings, 1970; Neal, 1997; Gilks and Wild, 1992).

For jointly updating $\boldsymbol{\delta}_i$, $\boldsymbol{\eta}_i^2$, $\theta_i$, $\sigma_i^2$ in Equation (33) at iteration $t$, a Metropolis-Hastings step is implemented. For this, the stage 1 samples are used for the proposal distribution, and the target distribution is the full hierarchical model (see Section 2.2 for complete details). Here, we randomly sample a candidate value $\{\boldsymbol{\delta}_i^{*(t)}, \boldsymbol{\eta}_i^{2*(t)}, \theta_i^{*(t)}, \sigma_i^{2*(t)}\}$ from the $A_i$ samples from stage 1 and label it $a_{it}$, where $\{\boldsymbol{\delta}_i^{*(t)}, \boldsymbol{\eta}_i^{2*(t)}, \theta_i^{*(t)}, \sigma_i^{2*(t)}\}$ has the following posterior density from Equation (30):

$$\{\boldsymbol{\delta}_i^{*(t)}, \boldsymbol{\eta}_i^{2*(t)}, \theta_i^{*(t)}, \sigma_i^{2*(t)}\} = \{\boldsymbol{\delta}_i^{(a_{it})}, \boldsymbol{\eta}_i^{2(a_{it})}, \theta_i^{(a_{it})}, \sigma_i^{2(a_{it})}\} \\ \sim p(\boldsymbol{y}_{ij}|\boldsymbol{\delta}_i, \boldsymbol{\eta}_i^2) p(\boldsymbol{\delta}_i|\theta_i, \sigma_i^2) p(\boldsymbol{\eta}_i^2) p(\theta_i) p(\sigma_i^2) = q(\boldsymbol{\delta}_i, \boldsymbol{\eta}_i^2, \theta_i, \sigma_i^2). \tag{34}$$

The target posterior density of the full model in Equation (33) is also assessed at the candidate value $\{\boldsymbol{\delta}_i^{*(t)}, \boldsymbol{\eta}_i^{2*(t)}, \theta_i^{*(t)}, \sigma_i^{2*(t)}\}$; this target posterior density includes priors for $\theta_i$ that are dependent on $\mu$ and $\tau^2$, as follows:

$$p(\boldsymbol{\delta}_i, \boldsymbol{\eta}_i^2, \theta_i, \sigma_i^2 | \mu, \tau^2, \boldsymbol{y}) \propto \\ \sim p(\boldsymbol{y}_{ij}|\boldsymbol{\delta}_i, \boldsymbol{\eta}_i^2) p(\boldsymbol{\delta}_i|\theta_i, \sigma_i^2) p(\boldsymbol{\eta}_i^2) p(\theta_i|\mu, \tau^2) p(\sigma_i^2) = h(\boldsymbol{\delta}_i, \boldsymbol{\eta}_i^2, \theta_i, \sigma_i^2). \tag{35}$$

The target-to-candidate ratio for the candidate value $\{\boldsymbol{\delta}_i^{*(t)}, \boldsymbol{\eta}_i^{2*(t)}, \theta_i^{*(t)}, \sigma_i^{2*(t)}\}$ is then

$$R\left(\delta_i^{*(t)}, \eta_i^{2*(t)}, \theta_i^{*(t)}, \sigma_i^{2*(t)}\right) = \frac{h\left(\delta_i^{*(t)}, \eta_i^{2*(t)}, \theta_i^{*(t)}, \sigma_i^{2*(t)}\right)}{q\left(\delta_i^{*(t)}, \eta_i^{2*(t)}, \theta_i^{*(t)}, \sigma_i^{2*(t)}\right)}$$

$$\propto \frac{p(y_{ij} | \delta_i^{(a_{it})}, \eta_i^{2(a_{it})}) p(\delta_i^{(a_{it})} | \theta_i^{(a_{it})}, \sigma_i^{2(a_{it})}) p(\eta_i^{2(a_{it})}) p(\theta_i^{(a_{it})} | \mu^{(t)}, \tau^{2(t)}) p\left(\sigma_i^{2(a_{it})}\right)}{p(y_{ij} | \delta_i^{(a_{it})}, \eta_i^{2(a_{it})}) p(\delta_i^{(a_{it})} | \theta_i^{(a_{it})}, \sigma_i^{2(a_{it})}) p(\eta_i^{2(a_{it})}) p(\theta_i^{(a_{it})}) p\left(\sigma_i^{2(a_{it})}\right)} \quad (36)$$

$$= \frac{p(\theta_i^{(a_{it})} | \mu^{(t)}, \tau^{2(t)})}{p(\theta_i^{(a_{it})})}.$$

Similarly, the target-to-candidate ratio for the previous value of the Markov chain $\{\delta_i^{(t-1)}, \eta_i^{2(t-1)}, \theta_i^{(t-1)}, \sigma_i^{2(t-1)}\}$ is

$$R\left(\delta_i^{(t-1)}, \eta_i^{2(t-1)}, \theta_i^{(t-1)}, \sigma_i^{2(t-1)}\right) = \frac{h\left(\delta_i^{(t-1)}, \eta_i^{2(t-1)}, \theta_i^{(t-1)}, \sigma_i^{2(t-1)}\right)}{q\left(\delta_i^{(t-1)}, \eta_i^{2(t-1)}, \theta_i^{(t-1)}, \sigma_i^{2(t-1)}\right)}$$

$$\propto \frac{p(y_{ij} | \delta_i^{(t-1)}, \eta_i^{2(t-1)}) p(\delta_i^{(t-1)} | \theta_i^{(t-1)}, \sigma_i^{2(t-1)}) p(\eta_i^{2(t-1)}) p(\theta_i^{(t-1)} | \mu^{(t)}, \tau^{2(t)}) p\left(\sigma_i^{2(t-1)}\right)}{p(y_{ij} | \delta_i^{(t-1)}, \eta_i^{2(t-1)}) p(\delta_i^{(t-1)} | \theta_i^{(t-1)}, \sigma_i^{2(t-1)}) p(\eta_i^{2(t-1)}) p(\theta_i^{(t-1)}) p(\sigma_i^{2(t-1)})} \quad (37)$$

$$= \frac{p(\theta_i^{(t-1)} | \mu^{(t)}, \tau^{2(t)})}{p(\theta_i^{(t-1)})}.$$

Thus, $r$ is given by the following, and the candidate value $\{\delta_i^{*(t)}, \eta_i^{2*(t)}, \theta_i^{*(t)}, \sigma_i^{2*(t)}\}$ will be accepted with the probability of minimum(1,$r$):

$$r = \frac{h(\delta_i^{(a_{it})}, \eta_i^{2(a_{it})}, \theta_i^{(a_{it})}, \sigma_i^{2(a_{it})})}{q(\delta_i^{(a_{it})}, \eta_i^{2(a_{it})}, \theta_i^{(a_{it})}, \sigma_i^{2(a_{it})})} \frac{q(\delta_i^{(t-1)}, \eta_i^{2(t-1)}, \theta_i^{(t-1)}, \sigma_i^{2(t-1)})}{h(\delta_i^{(t-1)}, \eta_i^{2(t-1)}, \theta_i^{(t-1)}, \sigma_i^{2(t-1)})}$$

$$= \frac{R\left(\delta_i^{(a_{it})}, \eta_i^{2(a_{it})}, \theta_i^{(a_{it})}, \sigma_i^{2(a_{it})}\right)}{R\left(\delta_i^{(t-1)}, \eta_i^{2(t-1)}, \theta_i^{(t-1)}, \sigma_i^{2(t-1)}\right)} = \frac{\frac{p(\theta_i^{(a_{it})} | \mu^{(t)}, \tau^{2(t)})}{p(\theta_i^{(a_{it})})}}{\frac{p(\theta_i^{(t-1)} | \mu^{(t)}, \tau^{2(t)})}{p(\theta_i^{(t-1)})}} \quad (38)$$

$$= \frac{p(\theta_i^{(a_{it})} | \mu^{(t)}, \tau^{2(t)})}{p(\theta_i^{(t-1)} | \mu^{(t)}, \tau^{2(t)})} \frac{p(\theta_i^{(t-1)})}{p(\theta_i^{(a_{it})})}.$$

Assuming effectively uniform priors in stage 1 for $\theta_i$, $r$ can be simplified further as

$$r = \frac{p(\theta_i^{(a_{it})} | \mu^{(t)}, \tau^{2(t)})}{p(\theta_i^{(t-1)} | \mu^{(t)}, \tau^{2(t)})} \frac{p(\theta_i^{(t-1)})}{p(\theta_i^{(a_{it})})} = \frac{p(\theta_i^{(a_{it})} | \mu^{(t)}, \tau^{2(t)})}{p(\theta_i^{(t-1)} | \mu^{(t)}, \tau^{2(t)})}. \quad (39)$$

Except for $\theta_i$, all distributional assumptions are the same in levels 1 to $c_s$ for both the stage 1 model and the full hierarchical model, and thus the $\delta_i, \eta_i^2$, and $\sigma_i^2$ parameters cancel in Equations (36) and (37). In addition, since data values are the same in the stage 1 model and the full hierarchical model, the likelihood terms also cancel in Equations (36) and (37). As a result, the ratio *r* for the Metropolis-Hastings step does not depend on the data sets, and stage 2 can be performed very rapidly. Note that the ratio *r* in Equation (39) is the same as for the three-level model in Section 2.2. The ratio *r* is again the ratio of priors for $\theta_i$ from the full model, evaluated at the candidate value for $\theta_i$ in the numerator, and the previous value of the Markov chain for $\theta_i$ in the denominator. Convergence of the MCMC sampler in stage 2 is determined using standard procedures (Cowles and Carlin [9]; Mengersen *et al*. [23]); we describe convergence diagnostics in Section 2.3.

Table 1. Minimum MCMC Efficiency.

| Data Set (a) | Number of Groups (b) | Two-Stage Analysis | | | Full Data Analysis (f) | MCMC Efficiency Improvement Factor, Two Stage Versus Full Data: (e)/(f)*0.50 |
| --- | --- | --- | --- | --- | --- | --- |
| | | Stage 1: Minimum of All Subsets (c) | Stage 2 (d) | Average Over Stage 1 and Stage 2 (e) | | |
| Simulation Study 1 | 20 | 9.18 | 63.44 | 36.31 | 0.59 | 30.8 |
| | 50 | 12.91 | 24.82 | 18.87 | 0.34 | 27.8 |
| | 100 | 15.56 | 12.29 | 13.93 | 0.18 | 38.7 |
| Simulation Study 2 | 20 | 1.74 | 11.39 | 6.57 | 0.15 | 21.9 |
| Real | 12 | 4.05 | 86.30 | 45.18 | 1.32 | 17.1 |

Note: Minimum MCMC efficiency [= minimum(Effective sample size / CPU time in seconds)] over all parameters for sampling 250,000 samples, after burnin. The MCMC efficiency improvement factor of two stage versus full data analysis is multiplied by 50%, since the two stage method requires twice as many samples in total versus the full data analysis. The R programming language computation times are based on computers with Linux Ubuntu 16.04 operating systems and Intel Xeon E3-1225 V2 CPU 3.2 GHz Processors with 16GB memory.

Table 2. Computation time (in minutes).

| Data Set | Number of Groups | Type of Time | Two-Stage Analysis | | | Total: Full Data Analysis | Two Stage Percent Time Reduction, After Burnin |
|---|---|---|---|---|---|---|---|
| | | | Stage 1: Maximum Over All Subsets | Stage 2 | Two Stage Total: Stage 1 plus Stage 2 | | |
| Simulation Study 1 | 20 | CPU | 88.40 (3.92) | 6.18 (0.25) | 94.58 (4.17) | 1,358.42 (53.92) | 93.0% |
| | | Elapsed | 88.43 (3.92) | 6.18 (0.26) | 94.61 (4.18) | 1,358.47 (54.04) | 93.0% |
| | 50 | CPU | 62.90 (2.19) | 15.71 (0.63) | 78.61 (2.82) | 2,386.27 (95.27) | 96.7% |
| | | Elapsed | 62.92 (2.19) | 15.72 (0.64) | 78.64 (2.83) | 2,386.47 (95.28) | 96.7% |
| | 100 | CPU | 52.24 (2.38) | 31.33 (1.26) | 83.57 (3.64) | 4,528.82 (181.03) | 98.2% |
| | | Elapsed | 52.25 (2.38) | 31.33 (1.26) | 83.58 (3.64) | 4,529.11 (181.03) | 98.2% |
| Simulation Study 2 | 20 | CPU | 301.14 (12.67) | 69.72 (2.79) | 370.86 (15.46) | 3,375.01 (138.39) | 89.0% |
| | | Elapsed | 301.15 (12.67) | 69.90 (2.80) | 371.05 (15.47) | 3,375.19 (138.39) | 89.0% |
| Real | 12 | CPU | 197.96 (8.05) | 5.86 (0.24) | 203.82 (8.29) | 586.62 (23.49) | 65.3% |
| | | Elapsed | 198.07 (8.05) | 5.87 (0.24) | 203.94 (8.29) | 586.64 (23.49) | 65.2% |

Note: R programming language computation time (in minutes) for sampling 250,000 samples for the simulation data sets and real data sets, after burnin. Burnin time (in minutes) for 10,000 samples is listed in parentheses. The computation times are based on computers with Linux Ubuntu 16.04 operating systems and Intel Xeon E3-1225 V2 CPU 3.2 GHz Processors with 16GB memory.

Table 3. Estimated relative $L_1$ and $L_2$ distances for the simulation data and real data; for parameters with indices, values are averaged over all indices.

| Data Set and Model | Number of Groups | Estimated Relative Distance | $\mu$ | $\tau^2$ | $\theta$ | $\sigma^2$ |
|---|---|---|---|---|---|---|
| Simulation Study 1, 3-Level Model | 20 | $L_1$ | 0.021 | 0.016 | 0.023 | 0.022 |
| | | $L_2$ | 0.023 | 0.014 | 0.024 | 0.024 |
| | 50 | $L_1$ | 0.021 | 0.023 | 0.023 | 0.023 |
| | | $L_2$ | 0.021 | 0.024 | 0.023 | 0.024 |
| | 100 | $L_1$ | 0.016 | 0.016 | 0.023 | 0.023 |
| | | $L_2$ | 0.018 | 0.018 | 0.023 | 0.024 |

| Data Set and Model | Number of Groups | Estimated Relative Distance | $\mu$ | $\Sigma$ | $\beta$ |
|---|---|---|---|---|---|
| Simulation Study 2, 4-Level Model | 20 | $L_1$ | 0.020 | 0.019 | 0.024 |
| | | $L_2$ | 0.020 | 0.018 | 0.025 |

| Data Set and Model | Number of Groups | Estimated Relative Distance | $\mu$ | $\tau^2$ | $\delta$ | $\eta$ | $\theta$ | $\sigma^2$ |
|---|---|---|---|---|---|---|---|---|
| Real Data, 4-Level Model | 12 | $L_1$ | 0.018 | 0.025 | 0.022 | 0.023 | 0.026 | 0.025 |
| | | $L_2$ | 0.017 | 0.023 | 0.022 | 0.023 | 0.024 | 0.031 |

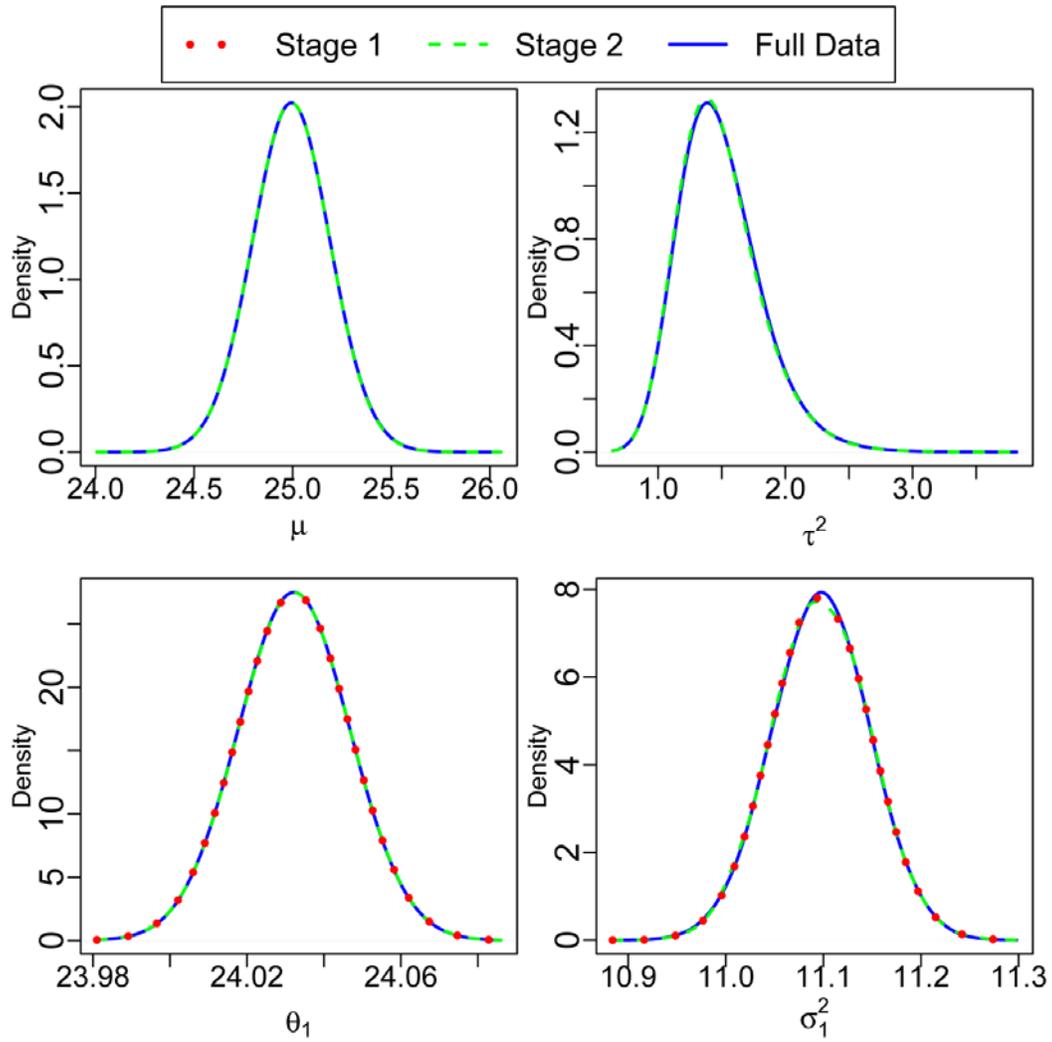

Figure 1. Marginal posterior distributions based on the stage 1, stage 2 and full data analysis for given parameters of the simulation study 1 data analysis, for $n = 50$ groups. Note that $\mu$ and $\tau^2$ are not estimated in stage 1.

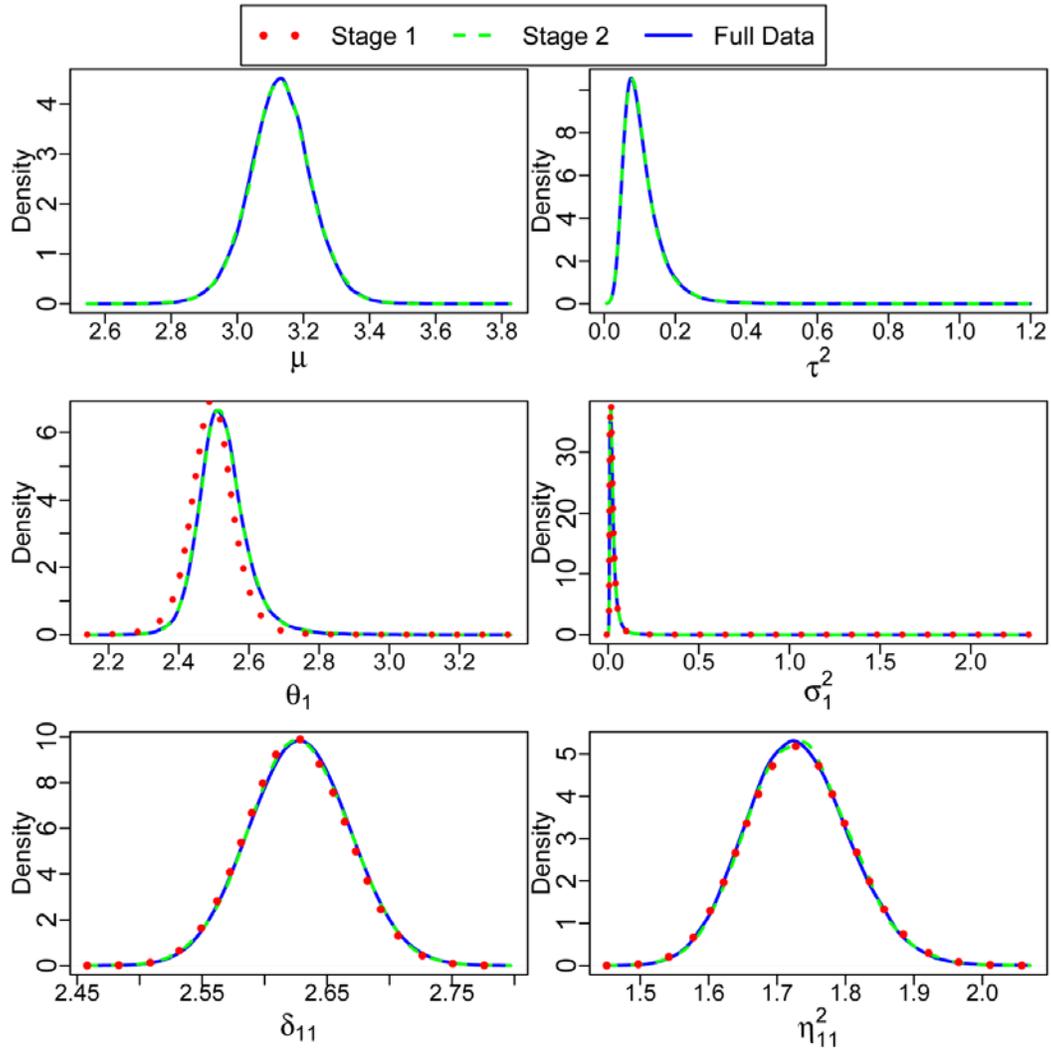

Figure 2. Marginal posterior distributions based on the stage 1, stage 2 and full data analysis for given parameters of the real data analysis. Note that $\mu$ and $\tau^2$ are not estimated in stage 1.